\documentclass[%
 reprint,
superscriptaddress,
nofootinbib,
 amsmath,amssymb,
 aps,prd,onecolumn
]{revtex4-2}
\usepackage{graphicx}
\usepackage{bm}
\usepackage{braket}
\usepackage{bbold}
\usepackage{amssymb}
\usepackage{amsmath,latexsym}
\usepackage{subfigure}
\usepackage{slashed}
\usepackage[T1]{fontenc}
\usepackage{multirow,array}
\usepackage{mathtools}
\usepackage{mathrsfs}
\usepackage{dsfont}
\usepackage{soul}
\usepackage[colorlinks=false,linktocpage=true]{hyperref}
\usepackage{hyperref}
\usepackage[utf8]{inputenc}
\usepackage[export]{adjustbox}
\usepackage{float}
\usepackage[makeroom]{cancel}
\usepackage{wrapfig}
\usepackage{ulem}
\usepackage{graphicx} 
\usepackage{amsmath}
\usepackage{physics}
\usepackage{color}
\usepackage{xcolor}
\usepackage{orcidlink}

\begin{document}

\newcommand*{\Bogazici}{Department of Physics, Bogazici University, Istanbul, Turkey}  
  
\newcommand*{\VT}{Department of Physics, Virginia Tech, Blacksburg, VA 24061, U.S.A.}


\title{\boldmath 
Resurgence of 
deformed genus-1 curves: A novel perturbative/nonperturbative relation}

\author{Atakan \c{C}avu\c{s}o\u{g}lu\,\orcidlink{0009-0005-5268-3266}}
\thanks{Current affiliation: University of Pennsylvania}
\email{atakanc@sas.upenn.edu}
\affiliation{Department of Physics and Astronomy, University of Pennsylvania, Philadelphia, PA 19104, U.S.A}
\affiliation{Department of Physics, Bo\u{g}azi\c{c}i University, Istanbul, Turkey}
\affiliation{Department of Electrical and Electronics Engineering, Bo\u{g}azi\c{c}i University, Istanbul, Turkey}

\author{Can Koz\c{c}az\,\orcidlink{0000-0002-7768-926X}}
\email{can.kozcaz@bogazici.edu.tr}
\affiliation{Department of Physics, Bo\u{g}azi\c{c}i University, Istanbul, Turkey}
\affiliation{Feza G\"{u}rsey Center for Physics and Mathematics, Bo\u{g}azi\c{c}i University, Istanbul, Turkey}
\affiliation{Niels Bohr Institute, Copenhagen University, Blegdamsvej 17, Copenhagen, 2100, Denmark}

\author{Kemal Tezgin\,\orcidlink{0000-0001-9492-9512}}
\email{kemaltezgin@vt.edu}
\affiliation{Department of Physics, Virginia Tech, Blacksburg, VA 24061, U.S.A.}

\preprint{APS/123-QED}

\begin{abstract}
We present a new perturbative/nonperturbative (P/NP) relation that applies to a broader class of genus-1 potentials, including those that possess real and complex instantons parametrized by a deformation parameter, such as polynomial and elliptic potentials. Our findings significantly extend the scope of quantum mechanical systems for which perturbation theory suffices to calculate the contributions of nonperturbative effects to energy levels, with or without the need for a boundary condition, depending on the potential. We further provide evidence for our results by predicting the corrections to the large-order behavior of the perturbative expansion for the Jacobi SD elliptic potential using the early terms of the instanton fluctuations that satisfy the P/NP relation.
\center{\it In memory of Durmu\c{s} Ali Demir}
\end{abstract}



\maketitle

\section{Introduction} 

It has long been recognized that an intricate relationship exists between perturbation theory and nonperturbative physics in certain quantum mechanical systems. The initial instances of such connections were first recognized between the early instanton fluctuation terms and large-order corrections in the perturbative expansion around the perturbative saddle of the path integral \cite{Bender:1969si,Bender:1973rz}. Such examples demonstrate the so-called resurgence phenomenon, which establishes concrete links between the asymptotically divergent perturbative expansions around perturbative and nonperturbative saddles, allowing one to unambiguously express an observable in a trans-series form \cite{Zinn-Justin:2004vcw, Zinn-Justin:2004qzw, Jentschura:2010zza}. 

Aside from the intricate early-late term relations between the monomials in the trans-series, \'{A}lvarez and collaborators further demonstrated the existence of a relationship between the early terms of the perturbative expansion of the perturbative saddle and the early terms of the fluctuations around nonperturbative instanton contributions; first for the cubic \cite{Alvarez:2000} and then for the double-well potentials \cite{Alvarez2004}. These were the first explicit examples of a new type of resurgence program since they offered a relationship between the low-order terms in the expansions around perturbative and nonperturbative saddles.

Later, Dunne and \"Unsal, from a different motivation and perspective, deepened these connections for several genus-1 \footnote{In this context, by the genus of a potential we mean the genus of the Riemann surface defined by the energy conservation relation $p^2= 2(E-V(x))$ in the complexified phase space $(x,p)$.} type of potentials \cite{Dunne:2013ada, Dunne:2014bca, Dunne:2016qix, Dunne:2016jsr}, and offered a constructive way to generate nonperturbative physics solely from perturbative results. For this class of potentials, they showed that \cite{Zinn-Justin:2004vcw, Zinn-Justin:2004qzw} two functions $B(E, g)$ and $A(E, g)$ satisfy a first-order differential equation when expressed in terms of $E(B, g)$ and $A(B, g)$:
\begin{equation}\label{classical_PNP}
    \frac{\partial E(B,g)}{\partial B}=-\frac{g}{S}\left(B+g\frac{\partial A(B,g)}{\partial g} \right) \, ,
\end{equation}
where $S$ is the instanton action and $g$ is the coupling constant. $B(E, g)$ and $A(E, g)$ respectively characterize the Rayleigh-Schr\"odinger perturbative expansion, and the single-instanton effects including fluctuations around the classical instanton solution. They were introduced in \cite{Zinn-Justin:2004qzw,Zinn-Justin:2004vcw} (see also the App.~\ref{App:Quantum-Periods} for a short review, and \cite{Cavusoglu:2023bai} for their computation from Picard-Fuchs equations) to express the exact quantization conditions for a set of quantum-mechanical systems, meaning that the energy levels $E_N$ can be calculated in a trans-series form by solving the quantization condition if $B(E,g)$ and $A(E,g)$ are known.

The computation of $B(E,g)$ is generally straightforward, as $E(B,g)$, the inverse of $B(E,g)$, is equal to the perturbation series in $g$ for the energy levels $E_N^{\text{(pert.)}}$ after setting $B = N+1/2$, which is the perturbative Bohr-Sommerfeld quantization condition. On the other hand, the instanton fluctuations $A(E,g)$ can be significantly challenging to compute even for very simple systems. Eq.~(\ref{classical_PNP}), called the P/NP (standing for ``perturbative/nonperturbative'') relation, avoids such complicated computations, and gives a very simple formula for generating the nonperturbative terms $A(E,g)$ directly from the perturbative expansion $B(E,g)$ to the same order in $g$.

However, we observe that Eq.~(\ref{classical_PNP}) holds for genus-1 potentials \textit{only} with vanishing residues at the poles of the classical momentum, $p(x) = \sqrt{2(E-V(x))}$, such as for the cubic \cite{Alvarez:2000, Gahramanov:2015yxk}, double-well \cite{Alvarez2004, Dunne:2013ada, Dunne:2014bca} or Sine-Gordon potentials \cite{Dunne:2013ada, Dunne:2014bca}. In these examples, the poles of $p$ are at infinity, and the vanishing of the residue can be due to a parity symmetry of $p$ (as in the double-well potential) or due to a branch cut of $p$ extending to the pole (as in the cubic potential). We refer to Refs.~\cite{Alvarez:2000, Alvarez2004, Dunne:2013ada, Dunne:2014bca, Dunne:2016jsr, Dunne:2016qix, Gorsky:2014lia, Gahramanov:2015yxk, Gahramanov:2016xjj, Basar:2015xna, Kozcaz:2016wvy, Codesido:2017dns, Basar:2017hpr, vanSpaendonck:2023znn} for various manifestations of the P/NP relation in the form of Eq.~(\ref{classical_PNP}).

In this work, our objective is to decipher the nonperturbative information from the perturbative one in a more unified way for a broader set of potentials. To this end, we propose a generalization of the P/NP relation to \textit{any} \footnote{However, we are still restricting ourselves to the case where $p$ has only one pole in its fundamental domain, in which case the moduli space of the Riemann surface can be locally parametrized in terms of three parameters $E$, $g$, and $\eta$.} one-parameter $\eta$ family of genus-1 potentials, $V(x;\eta)$, which we initiated in \cite{Cavusoglu:2023bai}. This generalization takes into account the possibility of a nonvanishing residue at the pole of $p$, in which case the residue $R(\eta)$ is a function of the deformation parameter $\eta$ which we introduce in the potential, such as one of the coefficients in a polynomial potential or the elliptic parameter for a doubly periodic potential. The novelty of our formula is not only in its encapsulating compact mathematical form but, more importantly, in the way it treats different types of known nonperturbative phenomena, such as real and complex (with any phase) instantons, on equal footing.

We argue that the P/NP relation for one-parameter families of genus-1 potentials takes the following generic form when the deformation of the potential is expressed in terms of the residue $R(\eta)$: 
\begin{equation} \label{parametricPNP_simplest}
  \boxed{ \partial_B E(B,\widetilde{R},g) = \frac{\partial_g \mathcal{A}(B,\widetilde{R},g)}{\partial_g \widetilde{S}(\widetilde{R},g)}}
\end{equation}
where $\mathcal{A}$ is the nonperturbative quantum period, which is closely related to the instanton function $A$ (cf. Eq.~(\ref{AandQuantumPeriod})), and $\widetilde {S}\equiv S/g$ and $\widetilde{R}\equiv R/g$ are the scaled instanton action and the scaled residue, respectively.

Alternatively, we will show that this equation can be expressed in the following equivalent form,
\begin{equation} \label{parametricPNP}
\frac{\partial E}{\partial B}= - \frac{g}{S(R)-R\dv*{S}{R}} \left( f(R)\, B + g \frac{\partial A}{\partial g} + R \frac{\partial A}{\partial R} \right)\,,
\end{equation}
where the connection of our P/NP relation to Eq.~(\ref{classical_PNP}) also becomes more transparent, as Eq.~(\ref{parametricPNP}) reduces to Eq.~(\ref{classical_PNP}) at $R=0$. However, it would be misleading to assume that the modification solely consists of making the instanton action deformation dependent; in fact, our proposal includes a derivative term with respect to the deformation parameter that is absent in Eq.~(\ref{classical_PNP}).
\\

\section{Dissecting the P/NP relation} 

The generalization from Eq.~(\ref{classical_PNP}) to Eqs.~(\ref{parametricPNP_simplest}-\ref{parametricPNP}) is itself a consequence of the genus-1 quantum topology of the systems we investigate. With the introduction of a deformation parameter, the systems in consideration generically have a nonzero residue, and the number of independent Wentzel–Kramers–Brillouin (WKB) periods (integrals of the quantum corrected momentum function over closed curves on the genus-1 torus, see the App.~\ref{App:Quantum-Periods} and \cite{Voros1983, Mironov:2009uv, Basar:2017hpr, Fischbach:2018yiu, Kreshchuk:2018qpf}) are increased to three: $B$, $\mathcal{A}$, and $\widetilde{R}_q$. By $\widetilde{R}_q$, we denote the residue of the WKB curve, or in other words, the residue of the classical momentum with the quantum corrections, which we will refer to as the ``quantum residue''.

Systems with nonvanishing residues can seemingly have multiple different nonperturbative $\mathcal{A}$ periods, such as different periods corresponding to the real and complex instantons (for which we will present an example), or the two different complex conjugate instantons (which we investigated in \cite{Cavusoglu:2023bai}). Still, the quantum topology of the systems reveals that seemingly different nonperturbative periods for genus-1 systems can only differ by a multiple of the quantum residue $\widetilde{R}_q$.

Therefore, it is natural to require the P/NP relation to hold for the various nonperturbative periods present in the system, which translates to an invariance of the P/NP relation under the shift of $\mathcal{A}$ by the quantum residue. For the systems in consideration, the quantum residue $\widetilde{R}_q$ is a function of the scaled residue $\widetilde{R}=R/g$ only. For poles at infinity, $\widetilde{R}_q = \widetilde{R}$ directly, which is the case for anharmonic polynomial potentials. On the other hand, for poles at finite points, the relation becomes
\begin{equation} \label{quantum_residue}
    \widetilde{R}_q = \sqrt{\widetilde{R}^2 - \frac{1}{4}} \, ,
\end{equation}
which is the case for doubly periodic potentials. With these considerations, it becomes clear why the Eqs.~(\ref{parametricPNP_simplest}-\ref{parametricPNP}) are the correct generalizations, as they are invariant under a shift of $A$ by the quantum residue $\widetilde{R}_q$ and $S$ by the unscaled residue $R$.

\begin{figure}
    \centering
    \includegraphics[width=7cm]{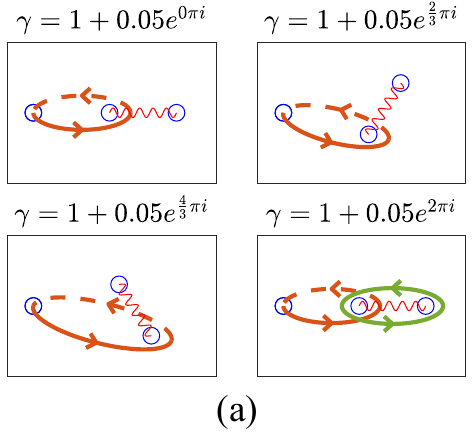} 
    \hspace{1cm}
    \includegraphics[width=7cm]{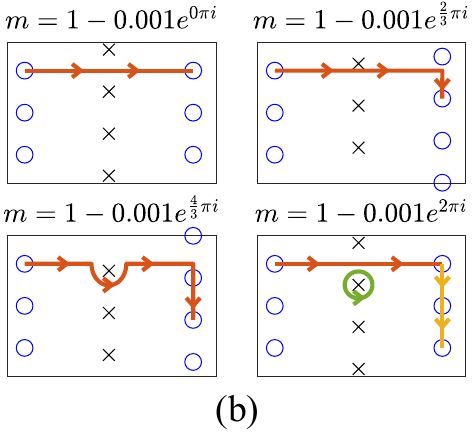}
    \caption{The deformation of the integration path of the instanton action on the Riemann surface of $\sqrt{2V}$ by the monodromy of the deformation parameters around the critical values, (a) for the deformed anharmonic potential (Eq.~(\ref{deformed_anharmonic_potential})), and (b) for the Jacobi SD elliptic potential (Eq.~(\ref{jacobi_sd_potential})). In both cases, the monodromy results in the addition of a residue term (green integration paths) to the instanton action.}
\label{fig:monodromic_deformations_small}
\end{figure}

This requirement on the P/NP relation can also be understood by considering the monodromies of the deformation parameters \cite{DDP}. As an example, consider the deformed anharmonic potential $V(x) = \frac{1}{2}x^2 - \gamma x^3 + \frac{1}{2}x^4$, which we previously investigated in \cite{Cavusoglu:2023bai}. Here, $\gamma$ is the deformation parameter we introduce, which cannot be removed by a scaling of $E$ or $g$. This potential reduces to the double-well potential for $\gamma=1$, for which a P/NP relation of the form in Eq.~(\ref{classical_PNP}) holds \cite{Alvarez2004, Dunne:2013ada}. However, for arbitrary values of $\gamma$, modifications to the P/NP relation are necessary to account for the nonvanishing residue. To understand the nature of this modification, notice that for values of $\gamma$ close to $1$, if $\gamma$ is rotated around $1$ by the deformation
\begin{equation}
    \gamma \rightarrow 1 + (\gamma-1)e^{2\pi i} \, ,
\end{equation}
the instanton action changes as
\begin{equation}
    S(\gamma) \rightarrow S(\gamma) + 2\pi R(\gamma) \, ,
\end{equation}
due to the crossing of a branch cut of the logarithm function appearing in the expression for the instanton action (cf. Eq.~(\ref{deformed_anharmonic_instanton_action})). The monodromic deformations of all such deformed genus-1 potentials, including the Jacobi elliptic potential in Eq.~(\ref{jacobi_sd_potential}), result in the addition of a residue term to the instanton action (see Figure (\ref{fig:monodromic_deformations_small})).

Inspired by these considerations, we propose that the P/NP relation applicable to deformed genus-1 potentials should also be of the linear form similar to Eq.~(\ref{classical_PNP}) and should be invariant under the monodromy of the potential. Therefore, proposing an Ansatz for the P/NP relation of the form
\begin{equation} \label{GenericPNP}
\frac{\partial E}{\partial B}= -g \left[ f(\eta)\,B + g(\eta)\, g \, \frac{\partial A}{\partial g} + h(\eta) \, \frac{\partial A}{\partial \eta} \right] \, ,
\end{equation}
one finds that to zeroth order in $g$, this equation becomes
\begin{equation}
    g(\eta) S(\eta) - h(\eta) \dv{S(\eta)}{\eta} = 1 \, .
\end{equation}
From the invariance under the monodromy ($S \rightarrow S + 2 \pi R$), one also finds that the coefficients $g(\eta)$ and $h(\eta)$ of this equation must satisfy
\begin{equation} \label{g_and_h_condition}
    g(\eta) R(\eta) - h(\eta) \dv{R(\eta)}{\eta} = 0 \, .
\end{equation}
From these two equations, one can solve for $g(\eta)$ and $h(\eta)$ and obtain Eq. (\ref{parametricPNP}) by changing variables from $\eta$ to $R$. 

This relation is also invariant under the addition of a residue term to the instanton function $A$ because the instanton function $A$ also has the same monodromy property as the instanton action. Under the monodromy of the deformation parameter, $A$ shifts by a term proportional to the quantum residue:
\begin{equation}
    A(g,B,\eta) \rightarrow A(g,B,\eta) + 2 \pi \widetilde{R}_q(g,\eta) \, .
\end{equation}

Nevertheless, such arguments are not sufficient to reveal the form of Eq.~(\ref{parametricPNP_simplest}), as it encapsulates more information than the Eq.~(\ref{parametricPNP}), which only relates the instanton function $A$ to the perturbative series $E$. Instead, the relation actually relates $E$ to the quantum period $\mathcal{A}$, the relation of which to the instanton function $A$ is \cite{Zinn-Justin:2004vcw,Cavusoglu:2023bai}
\begin{equation}\label{AandQuantumPeriod}
    \mathcal{A}(B,g,R) = A(B,g,R) + a_1(R) B + B \log(gB ) \\+ \sum_{k=2}^{\infty} \frac{B_k(1/2)}{k (k-1)} B^{1-k} \, ,
\end{equation}
where $B_k(x)$ are the Bernoulli polynomials. Here, $a_1(R)$ is an important term as it depends on $R$. It is the coefficient of the $E^1 g^0$ term in the series expansion of $\mathcal{A}$, and therefore is also the coefficient of the term linear in $B$, and is chosen such that $A$ has no term of order $g^0$. We find that replacing $A$ by $\mathcal{A}$ exactly cancels the function $f(R)$ appearing in Eq.~(\ref{parametricPNP}), which is a factor that is independent of the instanton function $A$ and is given by
\begin{equation} \label{f_function_PNP}
    f(R) = 1 + R \dv*{a_1(R)}{R} \, .
\end{equation}
Since the instanton function $A$ has no term of order $g^0$, $f(R)B$ is precisely proportional to the coefficient of the $g^1$ term in $\pdv*{E}{B}$. In practical calculations, it was previously thought that the $+B$ term in Eq.~(\ref{classical_PNP}) was there only to compensate for the nonexistent $g^0$ term in $A$, and that the $g^1$ term in $\pdv*{E}{B}$ did not affect $A$. But in the P/NP relation in Eq.~(\ref{parametricPNP_simplest}), it becomes clear that the $g^1$ term in $\pdv*{E}{B}$ in fact determines the $B^1 g^0$ term in the quantum period $\mathcal{A}$ through the relation for $a_1(R)$ given in Eq.~(\ref{f_function_PNP}), which we have verified for the potentials of our interest.

We stress that the P/NP relation stated in Eq.~(\ref{parametricPNP}) is a first-order linear partial differential equation and typically requires assigning an initial or boundary value to get a unique solution. Specifically, the solution $A(B,g,\eta)$ is unique only up to some function of the scaled residue, $F(R/g)$. Therefore, to $n$'th order in $g$, the solution $A(B,g,\eta)$ is unique only up to $R^{-n}$, the coefficient of which can be determined without any additional knowledge only when $A(B,g,\eta)$ is expected to be continuous at a critical value $\eta_c$ at which the residue $R(\eta)$ vanishes. We refer to the App.~\ref{App:PNP_general_solution} for the general solution of the P/NP relation.

If the P/NP relation can be solved uniquely, the nonperturbative function $A$ can be generated solely based on the perturbative function $B$. Conversely, if an initial or boundary value is necessary for the P/NP relationship, one needs to evaluate the nonperturbative function $A$ for a specific value of the deformation parameter, after which equation Eq.~(\ref{parametricPNP}) can be applied to extend the results to the entire range of the deformation parameter. We will illustrate both cases in the next section.
\\

\section{Examples} 

We provide two examples to illustrate the applicability of our P/NP relation in Eq.~(\ref{parametricPNP}). The first one is the deformed anharmonic potential
\begin{equation} \label{deformed_anharmonic_potential}
    V(x,\gamma) = \frac{1}{2}x^2-\gamma x^3+\frac{1}{2}x^4 \, .
\end{equation}
For this potential, the instanton action and the residue of $p$ at infinity read
\begin{align} \label{deformed_anharmonic_instanton_action}
    S(\gamma) &= -\frac{2}{3} + \gamma^2 + \frac{1}{2}\left(\gamma^2-1\right) \gamma \log(\frac{\gamma - 1}{\gamma + 1}) \, , \\
    \label{deformed_anharmonic_residue}
    R(\gamma) &= \frac{i}{2} \gamma\left(\gamma^2-1\right).
\end{align}
The second example is the doubly periodic Jacobi elliptic potential
\begin{equation} \label{jacobi_sd_potential}
    V_{\text{sd}^2}(x,m) = \frac{1}{2} \text{sd}^2(x,m) \, ,
\end{equation}
where $\text{sd}(x,m)$ is the Jacobi SD elliptic function \cite{NIST:DLMF_sd}.

The nonperturbative phenomena appearing in these two examples are, on the surface, completely different. For the deformed anharmonic potential, either real or complex instantons can be in effect depending on the value of the deformation parameter $\gamma$. For the elliptic potential, it is known \cite{Basar:2013eka} that, except at the extreme values of $m$, both real and ghost (complex instantons with negative-valued actions) instantons must be simultaneously taken into account, and they both simultaneously govern the large-order behavior of the perturbative expansion for the energy levels. Although these two phenomena are seemingly unrelated, the genus-1 topology of both systems allows them to be united under a common P/NP formula, Eq.~(\ref{parametricPNP}).

We have already investigated the P/NP relation and the large-order corrections for the deformed anharmonic potential in Eq.~(\ref{deformed_anharmonic_potential}) in an earlier work \cite{Cavusoglu:2023bai}. We showed that the P/NP relation for this potential must be modified as
\begin{equation} \label{gammaPNP}
    \frac{\partial E}{\partial B}=-\frac32 g \left[ \left(5\gamma^2-1\right)B + \left(3\gamma^2-1\right)g\frac{\partial A}{\partial g}  + \left(\gamma^2-1\right)\gamma \frac{\partial A}{\partial \gamma} \right] \, .
\end{equation}
In light of the P/NP relation in Eq.~(\ref{parametricPNP}) and the discussion in the previous section, we now understand that Eq.~(\ref{gammaPNP}) can easily be derived from Eq.~(\ref{parametricPNP}) by a simple change of variables and noting the instanton action and the residue in Eqs.~(\ref{deformed_anharmonic_instanton_action}-\ref{deformed_anharmonic_residue}). For this potential, the boundary condition for solving the P/NP relation can be fixed uniquely by considering the critical values of the deformation parameter, which was explained in the appendix of \cite{Cavusoglu:2023bai}.

The P/NP relation for the elliptic potential can also be similarly derived. For this potential, the instanton action and the residue of $p$ at the pole at $x = K(m) + i K(1-m)$ read
\begin{equation}
    S_{\text{sd}^2}(m) = \frac{2 \sin^{-1}(\sqrt{m})}{\sqrt{(1-m)m}}\,, \quad R_{\text{sd}^2}(m) = \frac{1}{\sqrt{(1-m)m}} \, .
\end{equation}
Based on Eq.~(\ref{parametricPNP}), then, the P/NP relation for $V_{\text{sd}^2}(x;m)$ can be written as
\begin{equation} \label{PNP-SD}
\frac{\partial E_{\text{sd}^2}}{\partial B}= -g \left[ \left(\frac12 - m\right) B + \left(\frac12 - m\right) g \frac{\partial A_{\text{sd}^2}}{\partial g} - \left(1-m \right) m \frac{\partial A_{\text{sd}^2}}{\partial m} \right] \, ,
\end{equation}
which reproduces the P/NP relation for Sine-Gordon \cite{Dunne:2013ada, Dunne:2014bca} at $m=0$ and the P/NP relation of the Sinh-Gordon potential at $m=1$ \footnote{It should be noted that this reduction is not apparent in the expression for the residue, which diverges at $m=0$ or $m=1$. The reason is that the doubly periodic elliptic potential becomes singly periodic at $m=0$ or $m=1$, and its pole structure changes. Therefore, the expression for the residue $R_{\text{sd}^2}(m)$ is not valid for $m=0$ or $m=1$.}. Nevertheless, it should be noted that the boundary condition necessary for solving for $A$ cannot be obtained by considering the critical values $m=0$ or $m=1$ where $R^{-n} = 0$. We refer to the App.~\ref{App:PNP_general_solution} for details.

Both examples demonstrate the remarkable link between the perturbative expansion around the perturbative saddle and the instanton function $A$, which encodes the single instanton contribution with fluctuations around them. More precisely, the local information generated around the perturbative saddle through perturbation theory together with a boundary condition is sufficient to construct the nonperturbative instanton function $A$ both for the deformed anharmonic potential and the Jacobi elliptic potential, regardless of the value of the deformation parameter. 

Additionally, we demonstrate in the App.~\ref{App:Lam\'{e}} that the Lam\'{e} potential \cite{NIST:DLMF_sn} also satisfies our P/NP relation.
\\

\section{Large-order corrections for the elliptic potential} 

It is well-known in the literature that perturbation theory is divergent and grows factorially in the presence of nonperturbative contributions \cite{Bender:1969si, Bender:1973rz}. In this section, we will validate our P/NP relation given in Eq.~(\ref{PNP-SD}) by investigating the large-order behavior of the perturbative series for the energy levels of the Jacobi SD potential.

The perturbative energy $E$ and the instanton function $A$ for this potential are calculated as
\begin{widetext}
\begin{align}
    E_{\text{sd}^2} &= B - g \left(\frac{1}{2} -m\right) \left[\frac18 + \frac12 B^2\right] - g^2 \left[B \left(\frac{3}{64}+\frac{m}{8}-\frac{m^2}{8}\right) + \frac{1}{16}B^3\right] - g^3 \left(\frac12 - m \right)\left[ \frac{9}{1024} + \frac{17}{128} B^2 + \frac{5}{64} B^4 \right] + \dots \, ,\\
    A_{\text{sd}^2} &= \frac{S_{\text{sd}^2}}{g} \left[1 - \frac18 \frac{g^2}{R_{\text{sd}^2}^2} + \dots\right] +g \left(\frac12 -m \right) \left[\frac{3}{16} + \frac34 B^2 \right] + g^2\left[\frac{17}{128}B + \frac{5}{32} B^3 - R_{\text{sd}^2}^{-2} \left(\frac{5}{96}B + \frac{7}{24}B^3\right)\right] + \dots \, ,
\end{align}
\end{widetext}
which we obtained by solving the Picard-Fuchs equation and calculating the quantum corrections \cite{Mironov:2009uv,Fischbach:2018yiu,Kreshchuk:2018qpf,Cavusoglu:2023bai}. It can easily be checked that these functions satisfy the P/NP relation in Eq.~(\ref{PNP-SD}).

To leading order, the prediction of resurgence theory for the large-order behavior of the perturbative series for the energy levels was investigated in \cite{Basar:2013eka}. It was shown that the self-duality property of the Jacobi SD function
\begin{equation}
    \text{sd}(x,m) = i \, \text{sd}(ix,1-m)
\end{equation}
enforces one to include the effects of both real and ghost instantons in the prediction for the large-order growth of the energy perturbative series because the self-duality mapping swaps the real and ghost instantons. For the Jacobi SD elliptic potential (\ref{jacobi_sd_potential}), the ghost instanton action is given by
\begin{equation}
    S_{\mathcal{G}}(m) = -S_{\text{sd}^2}(1-m) = -\frac{2 \sin^{-1}(\sqrt{1-m})}{\sqrt{(1-m)m}} \, ,
\end{equation}
and the instanton/anti-instanton actions are given by
\begin{equation}
    S_{\mathcal{I}\bar{\mathcal{I}}}(m) = 2 S_{\text{sd}^2}(m)\, , \quad S_{\mathcal{G}\bar{\mathcal{G}}}(m) = 2 S_{\mathcal{G}}(m)\, .
\end{equation}

We recognize that these two actions differ by a multiple of the residue:
\begin{equation}
    S_{\mathcal{G}\bar{\mathcal{G}}}(m) = S_{\mathcal{I}\bar{\mathcal{I}}}(m) - 2\pi R_{\text{sd}^2}(m) \, .
\end{equation}
The reason is that these two instanton/anti-instanton actions are contour integrals on the Riemann surface of $\sqrt{2V}$, which is effectively genus-0. Therefore, the only nonzero period of $\sqrt{2V}$ is the residue, and contour integrals with common start and end points can only differ by the residue.

The self-duality of the Jacobi SD function manifests itself as the invariance of the perturbative series of the energy and the P/NP relation under the mappings $m \rightarrow 1-m$ and $g \rightarrow -g$:
\begin{equation}
    E_{\text{sd}^2}(B, g, m) = E_{\text{sd}^2}(B, -g, 1-m) \, ,
\end{equation}
but the instanton function $A_{\text{sd}^2}$ changes under the same mapping. The reason is that the instanton function depends on whether the period in consideration behind it is the real or the ghost period corresponding to the real or ghost instanton. Therefore, we actually have two different instanton functions, which we will call the real and ghost instanton functions.

We define
\begin{align}
    A_{\mathcal{I}\bar{\mathcal{I}}}(B,g,m) &= 2A_{\text{sd}^2}(B,g,m) \, , \label{A_instanton-antiinstanton} \\ A_{\mathcal{G}\bar{\mathcal{G}}}(B,g,m) &= 2A_{\text{sd}^2}(B,-g,1-m) \label{A_ghost-antighost}
\end{align}
as the real and ghost instanton/anti-instanton functions. These two functions correspond to periods on the WKB curve and can only differ by the quantum residue of the WKB curve:
\begin{equation} \label{A_shift_by_quantum_res}
    A_{\mathcal{G}\bar{\mathcal{G}}}(B,g,m) = A_{\mathcal{I}\bar{\mathcal{I}}}(B,g,m) - 2\pi \widetilde{R}_{q}(g,m) \, ,
\end{equation}
where the quantum residue is given by Eq.~(\ref{quantum_residue}). Since the instanton function is expressed as a series in $g$, the quantum residue should also be expanded into a series
\begin{equation}
    \widetilde{R}_q = \frac{R}{g}\left[1 - \frac{g^2}{8 R^2} - \frac{g^4}{128 R^4} - \frac{g^6}{1024R^6} + \dots\right] \, .
\end{equation}

Due to these considerations, having two different $A$ functions does not invalidate the P/NP relation. Since the two functions differ only by a term proportional to the quantum residue, which is a function of the scaled residue $\widetilde{R} = R/g$ only, they both satisfy the P/NP relation because the P/NP relation is invariant under the shift of $A$ by the quantum residue, as explained before.

\begin{figure}
    \centering
    \includegraphics[width=8cm]{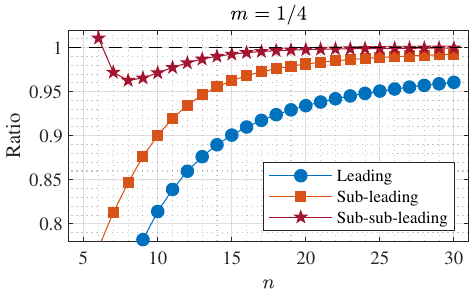} \hspace{1cm}
    \includegraphics[width=8cm]{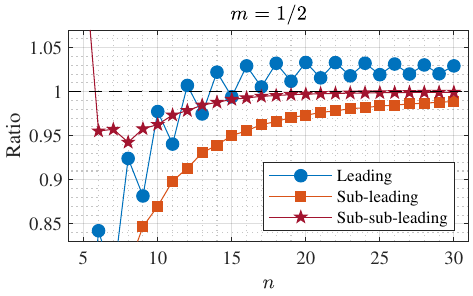}
    \caption{The ratios $a_n^{\text{(exact)}}/a_n^{\text{(growth)}}$ of the exact coefficients $a_n$ of the perturbative expansion and the large-order growth of the perturbative coefficients, Eq.~(\ref{SD2:large_order_prediction}), in the ground state for the Jacobi SD potential, $V = 1/2\, \text{sd}^2(x,m)$, plotted for $m=1/4$ and $1/2$. The ratios are plotted for the leading (Eq.~(\ref{eq:large_order_leading})), sub-leading (up to $b_1$ in Eq.~(\ref{SD2:large_order_prediction})), and sub-sub-leading (up to $b_2$ in Eq.~(\ref{SD2:large_order_prediction})) corrections to $a_n^{\text{(growth)}}$. For $m=1/2$, the odd coefficients in $n$ vanish, but the limit of the ratios as $m\rightarrow1/2$ exists, which are the values plotted for odd $n$. These results extend symmetrically to $m>1/2$ due to the self-duality of the potential.}
    \label{Fig:large-order-corrections-SD2}
\end{figure}

From self-duality considerations, the large orders of the perturbative series for the ground state energy $E_0^{\text{(pert.)}}=\sum_{n=0}a_n g^n$ grow to leading order as \cite{Basar:2013eka}
\begin{equation} \label{eq:large_order_leading}
    a_n^\text{(leading)} = -\frac{8n!}{\pi}  \left(\frac{1}{(S_{\mathcal{I}\bar{\mathcal{I}}})^{n+1}} - \frac{1}{(S_{\mathcal{G}\bar{\mathcal{G}}})^{n+1}} \right) \, .
\end{equation}

This large-order prediction is only up to the leading order, whereas the sub-leading orders of the large-order behavior are governed by the low orders of the instanton function $A$. In particular, they are directly related to the coefficients of the so-called fluctuation factor. In our case, we have two different fluctuation factors corresponding to the real and ghost instantons. The real fluctuation factor is calculated with the real instanton/anti-instanton action and function
\begin{equation}
    F_{\mathcal{I}\bar{\mathcal{I}}} = \exp\left[-A_{\mathcal{I}\bar{\mathcal{I}}}(B,g,m) + \frac{S_{\mathcal{I}\bar{\mathcal{I}}}(m)}{g}\right] \times \frac{\partial E(B,g,m)}{\partial B} \, ,
\end{equation}
and the ghost fluctuation factor can be calculated by taking the dual of the real fluctuation factor:
\begin{equation}
    F_{\mathcal{G}\bar{\mathcal{G}}}(B, g,m) = F_{\mathcal{I}\bar{\mathcal{I}}}(B, -g,1-m) \, .
\end{equation}
Then, the large-order prediction with corrections coming from the fluctuation factors is given by
\begin{equation}\label{SD2:large_order_prediction}
    a_n = - \frac{8n!}{\pi} \left[\frac{1}{(S_{\mathcal{I}\bar{\mathcal{I}}})^{n+1}}\left(1+\frac{b^{\mathcal{I}\bar{\mathcal{I}}}_1\,S_{\mathcal{I}\bar{\mathcal{I}}}}{n}+\frac{b^{\mathcal{I}\bar{\mathcal{I}}}_2\,S_{\mathcal{I}\bar{\mathcal{I}}}^2}{n (n-1)}+\dots\right)
    -  \frac{1}{(S_{\mathcal{G}\bar{\mathcal{G}}})^{n+1}} \left(1+\frac{b^{\mathcal{G}\bar{\mathcal{G}}}_1\,S_{\mathcal{G}\bar{\mathcal{G}}}}{n}+\frac{b^{\mathcal{G}\bar{\mathcal{G}}}_2\,S_{\mathcal{G}\bar{\mathcal{G}}}^2}{n (n-1)}+\dots\right) \right] \, ,
\end{equation}
where the coefficients $b^{\mathcal{I}\bar{\mathcal{I}}}_n$ and $b^{\mathcal{G}\bar{\mathcal{G}}}_n$ represent the coefficients of $g^n$ in $F_{\mathcal{I}\bar{\mathcal{I}}}(B,g,m)$ and $F_{\mathcal{G}\bar{\mathcal{G}}}(B,g,m)$, respectively, for the ground state $B=1/2$. 

In Figure (\ref{Fig:large-order-corrections-SD2}), we analyze the ratio between the exact coefficients $a_n$ calculated with \cite{Sulejmanpasic:2016fwr} and the predicted outcomes from Eq.~(\ref{SD2:large_order_prediction}) in the ground state, which shows the impact of the fluctuation terms on the large-order factorial growth of the perturbative series. The significance of including ghost instantons in the large-order growth of the perturbative series was shown in \cite{Basar:2013eka} only to leading order. Here, we demonstrate for the first time that the prediction of large-order terms can be improved further by including the fluctuation terms.

The Jacobi SD elliptic potential considered here represents a concrete expression of the idea of resurgence: the perturbative expansion around the perturbative saddle encodes a wealth of information about the nonperturbative saddles. Complex or negatively valued instanton actions should not be dismissed in the analysis of semiclassical expansion as they ultimately play a role in the large-order growth of the perturbative saddle. Incorporating the nonzero residues made this realization possible to a greater extent for the elliptic potentials, and any other genus-1 potential with a deformation parameter, which was not taken into account in the earlier considerations of the P/NP relation.
\\

\begin{acknowledgments}
We thank Mithat \"{U}nsal for stimulating discussions. CK would like to thank the warm hospitality of NBI. The work of CK was supported by the Scientific and Technological Research Council of Turkey (T\"{U}B\.ITAK) under the Grant Numbers 120F184 and 220N106. CK thanks T\"{U}B\.ITAK for their support. This article is based upon work from COST Action 21109 CaLISTA and COST Action 22113 THEORY-CHALLENGES, supported by COST (European Cooperation in Science and Technology).
\end{acknowledgments}

\appendix

\section{WKB and quantum periods}\label{App:Quantum-Periods}

The Schr\"odinger equation for the perturbed potential (setting $m = \hbar = 1$) is
\begin{equation} \label{WKBSchrodinger}
    \left[-\frac{1}{2}\pdv[2]{x} + \frac{1}{g}V(\sqrt{g}x)\right] \psi(x) = E \psi(x) \, .
\end{equation}
The potential $V(x)$ is chosen such that $\lim_{g\rightarrow0} V(\sqrt{g} x)/g = x^2/2$, which means that this equation describes a perturbation of the harmonic oscillator with perturbation parameter $g$. If one changes variables to $z = \sqrt{g} x$, the Schr\"odinger equation becomes
\begin{equation}
    \left[-\frac{g^2}{2} \pdv[2]{z} + V(z) \right] \psi(z) = \xi \psi(z) \, ,
\end{equation}
where $\xi = g E$, and $g$ takes on the role of $\hbar$ in the Schr\"odinger equation. WKB analysis proceeds by substituting $\psi(z) = \exp(\frac{i}{g} \int_{z_0}^z \rho(z^\prime) \dd{z^\prime})$, after which one obtains a Riccati equation
\begin{equation} \label{Riccati_eqn}
    \rho(z; \xi, g)^2 - i g \pdv{\rho(z;\xi,g)}{z} = 2(\xi - V(z)) \,.
\end{equation}
Here, one regards $\xi$ as a continuous variable instead of trying to solve for the energy eigenvalues, and this equation reduces to the classical energy conservation equation at $g=0$. For this reason, $\rho$ can be regarded as a ``quantum corrected'' version of the classical momentum $p$.

One important property of this equation is that if $\rho(z;\xi,g)$ is a solution, then $-\rho(z;\xi,-g)$ is also a solution. Therefore, the parts of $\rho(z;\xi,g)$ even and odd in $g$ are related to each other as
\begin{equation} \label{relation_rhoodd_rhoeven}
    \rho_{\text{odd}}(z;\xi,g) = \frac{i g}{2} \pdv{z} \log(\rho_{\text{even}}(z;\xi,g)) \, ,
\end{equation}
where $\rho_{\text{odd}}(z;\xi,g) = \frac12 \left(\rho(z;\xi,g) - \rho(z;\xi,-g)\right)$ and $\rho_{\text{even}}(z;\xi,g) = \frac12 \left(\rho(z;\xi,g) + \rho(z;\xi,-g)\right)$. For this reason, specifying $\rho_{\text{even}}$ is enough to specify $\rho$ completely. This relation can be derived by substituting the two different solutions $\rho_{\text{odd}} \pm \rho_{\text{even}}$ into the Riccati equation, and taking the difference.

The Riccati equation is solved formally as power series in $g$ by substituting $\rho(z; \xi, g) = \sum_{n=0}^\infty \rho_n(z;\xi) g^n$, where $\rho_0 = \pm \sqrt{2(\xi - V(z))}$. One obtains a recursion relation for the coefficients $\rho_n$:
\begin{equation}
    \rho_n = \frac{1}{2\rho_0} \left(i \pdv{\rho_{n-1}}{z} - \sum_{l=1}^{n-1}\rho_l \rho_{n-l} \right) \, .
\end{equation}

The WKB periods or the ``quantum periods'' are defined as the integrals of $\rho_{\text{even}}(z;\xi,g)$ over closed loops on the Riemann surface defined by the classical momentum:
\begin{equation}
    \sigma_{\mathcal{C}}(\xi,g) \equiv \oint_{\mathcal{C}} \rho_{\text{even}}(z;\xi,g) \dd{z} \, .
\end{equation}
This integral is understood to be evaluated term by term in the expansion coefficients $\rho_n$.

To switch back to $(x,E)$ from $(z,\xi)$, one needs to scale $\sigma\rightarrow \sigma/g$, which gives the quantum periods for the potential $V(\sqrt{g}x)/g$. With this notion, the perturbative quantum period $B(E,g)$ that appears in the P/NP relation can be defined as
\begin{equation}
    B(E,g) \equiv \frac{1}{2\pi g} \sigma_{\mathcal{C}_{\text{pert.}}}(g E, g) \, ,
\end{equation}
where $\mathcal{C}_{\text{pert.}}$ is the perturbative cycle that loops around the two turning points of the classical momentum $p = \sqrt{2(\xi - V(x))}$ around the minimum at which the perturbative expansion is performed. The nonperturbative quantum period $\mathcal{A}(E,g)$ can be defined as
\begin{equation}
    \mathcal{A}(E,g) \equiv \frac{1}{i g} \sigma_{\mathcal{C}_{\text{nonpert.}}}(g E, g) \, ,
\end{equation}
where $\mathcal{C}_{\text{nonpert.}}$ is the nonperturbative cycle that loops around the two turning points between which the particle tunnels. The instanton function $A(E,g)$ is closely related to the nonperturbative quantum period, the precise relationship being
\begin{equation}
    \mathcal{A}(B,g,R) = A(B,g,R) + a_1(R) B + B \log(gB ) + \sum_{k=2}^{\infty} \frac{B_k(1/2)}{k (k-1)} B^{1-k} \, ,
\end{equation}
as conjectured in \cite{Zinn-Justin:2004vcw}. $B(E,g)$ and $A(E,g)$ were introduced in \cite{Zinn-Justin:2004qzw,Zinn-Justin:2004vcw} to define the quantization conditions for a set of quantum-mechanical systems.

We define the ``quantum residue'' $R_q$ as the residue of $\rho_{\text{even}}$ at the pole inherited from the pole of $p = \sqrt{2(\xi - V(z))}$. If $p$ has a pole of order 1 at a finite point $z_0$ with residue $R$, its series expansion is
\begin{equation}
    p = \frac{R}{z-z_0} + \sum_{k=0}^{\infty} c_k (z-z_0)^k \, .
\end{equation}
Similarly, the series expansion of $\rho_{\text{even}}$ would be of the form
\begin{equation}
    \rho_{\text{even}} = \frac{R_q}{z-z_0} + \sum_{k=0}^{\infty} b_k (z-z_0)^k \, .
\end{equation}
Then, relating $\rho_{\text{odd}}$ to $\rho_{\text{even}}$ using Eq.~(\ref{relation_rhoodd_rhoeven}), one finds that the series expansion of $\rho = \rho_{\text{even}} + \rho_{\text{odd}}$ is
\begin{equation}
    \rho = \frac{R_q - \frac{ig}{2}}{z-z_0} + \sum_{k=0}^{\infty} b^{\prime}_k (z-z_0)^k \, ,
\end{equation}
and substituting into Eq.~(\ref{Riccati_eqn}), one finds that to order $(z-z_0)^{-2}$, we have
\begin{equation}
    R_q^2 + \frac{g^2}{4} = R^2 \, ,
\end{equation}
and in terms of the scaled residues $\widetilde{R} = R/g$ and $\widetilde{R}_q = R_q/g$, we find
\begin{equation}
    \widetilde{R}_q = \pm \sqrt{\widetilde{R}^2 - \frac{1}{4}} \, .
\end{equation}
If the same calculation is repeated for higher order poles, one finds that $\widetilde{R}_q=\pm R$, which is also the case for poles at infinity in most situations.
 
One application of the quantum periods is the perturbative Bohr-Sommerfeld quantization condition, which is
\begin{equation}
    \sigma_{\mathcal{C}_{\text{pert.}}}(\xi, \hbar) = 2 \pi \hbar \left(N + \frac{1}{2}\right) \, ,
\end{equation}
which, when written in terms of $B(E,g)$ becomes $B(E,g) = N + 1/2$. Therefore, if one inverts $B(E,g)$ to obtain $E(B,g)$, substituting $B = N + 1/2$ gives the standard Rayleigh-Schr\"odinger perturbation series $E_N^{\text{(pert.)}} = \sum_{n=0} a_{N,n}g^n$ for the energy levels, where $N$ denotes the $N$'th excited state.

The exact quantization condition also takes a very similar form to the perturbative Bohr-Sommerfeld quantization condition: 
\begin{equation}
    \sigma_{\mathcal{C}}(\xi,\hbar) = 2\pi \hbar \left(N+\frac{1}{2}\right) \, ,
\end{equation}
but $\sigma_{\mathcal{C}}(\xi,\hbar)$ is evaluated for a different contour $\mathcal{C}$, which has to enclose all zeros of the wave function. $\sigma_{\mathcal{C}}(\xi,\hbar)$ is in general a linear combination of the perturbative period $B$, nonperturbative period $\mathcal{A}$, and the residue $R$ for genus-1 potentials. Therefore, the exact quantization condition can be expressed in terms of $A(E,g)$ and $B(E,g)$ \cite{Zinn-Justin:2004vcw,Cavusoglu:2023bai}.

\section{General solution of the P/NP relation} \label{App:PNP_general_solution}

Here, we explain how to solve the P/NP relation in the form of Eq.~(\ref{GenericPNP}), given the information of $E(B,g,\eta)$. It is already known that $A(B,g,\eta)$ must be of the form
\begin{equation}
    A(B,g,\eta) = \frac{S(\eta)}{g} + \sum_{n=1}^{\infty} g^n \phi_n(B,\eta),
\end{equation}
and substituting this into Eq.~(\ref{GenericPNP}), to order $g^{n+1}$, we find
\begin{equation}
    n \, g(\eta) \, \phi_n(B,\eta) + h(\eta) \pdv{\phi_n(B,\eta)}{\eta} = - \psi_{n+1}(B,\eta),
\end{equation}
where $\psi_{n}(B,\eta)$ is the coefficient of the $g^n$ term in $\pdv*{E}{B}$. This equation is an inhomogeneous linear differential equation. Using Eq.~(\ref{g_and_h_condition}), it can be easily checked that the homogeneous solution is
\begin{equation}
    \phi_n^{\text{(hom.)}}(\eta) = R(\eta)^{-n}.
\end{equation}
The inhomogeneous solution can be found by rewriting the equation in terms of $\varphi_n(B,\eta) = R(\eta)^n \phi_n(B,\eta)$ as
\begin{equation}
    h(\eta) R(\eta)^{-n} \pdv{\varphi_n(B,\eta)}{\eta} = - \psi_{n+1}(B,\eta),
\end{equation}
and integrating:
\begin{equation}
    \phi_n^{\text{(inhom.)}}(B,\eta) = - R(\eta)^{-n}\int\frac{R(\eta)^n}{h(\eta)} \psi_{n+1}(B,\eta) \dd{\eta}.
\end{equation}
Therefore, the general solution for the coefficients $\phi_n(B,\eta)$ is
\begin{equation}
    \phi_n(B,\eta) = \phi_n^{\text{(inhom.)}}(B,\eta) + c_n(B) R(\eta)^{-n}.
\end{equation}

To have a complete solution for $A(B,g,\eta)$ solely from the knowledge of $E(B,g,\eta)$, $R(\eta)$, and $S(\eta)$, the coefficients of the homogeneous solution, $c_n(B)$, must be determined. The standard method of determining these coefficients is to provide a boundary condition, specifically, it is necessary to know the solution for $A(B,g,\eta)$ for one specific value of the deformation parameter $\eta$, for which the residue must be finite to have nonzero $R^{-n}$.

In the case of the deformed anharmonic potential in Eq.~(\ref{deformed_anharmonic_potential}), for which we gave the solution in \cite{Cavusoglu:2023bai}, considering the instanton function $A$ for $\gamma=0$ led us to the boundary condition necessary for solving $A$ for any value of the deformation parameter $\gamma$. Since $R(\gamma = 0)=0$, $R^{-n}$ blows up at $\gamma = 0$. Furthermore, since $\gamma=0$ is not a critical value for the genus-1 curve corresponding to the classical momentum (i.e., the topology of the Riemann surface does not change at $\gamma = 0$ when $E=0$), $A(B,g,\gamma)$ must be continuous at $\gamma=0$. From both considerations, it was found that $c_n(B) = 0$ for all $n$ such that $A(B,g,\gamma)$ has a finite expression at $\gamma=0$.

On the other hand, such considerations are not useful in the solution of $A(B,g,m)$ for the Jacobi SD elliptic potential in Eq.~(\ref{jacobi_sd_potential}). For this potential, the only specific values of $m$ that we can use are $m=0$ and $m=1$, but for both values of $m$, $R(m)^{-n} = 0$ for $n \geq 1$. Therefore, it is not possible to determine the coefficients $c_n(B)$ by having the information of $A$ at $m=0$ and $m=1$. Self-duality considerations,
\begin{equation}
    A_{\text{sd}^2}(B,-g,1-m) = A_{\text{sd}^2}(B,g,m) - \pi \widetilde{R}_{q}(g,m) \, ,
\end{equation}
can be applied to determine $c_n(B)$ for odd $n$, but the coefficients for even values of $n$ remain undetermined.

\section{P/NP relation for the Lam\'{e} equation}\label{App:Lam\'{e}}

The Lam\'{e} equation written in the Jacobian form is given by \cite{NIST:DLMF_sn}
\begin{equation}
    \dv[2]{w}{z} + \left( h - \nu \left( \nu+ 1\right) m \, \text{sn}^2 (z,m) \right) w = 0,
\end{equation}
which can be converted to a Schr\"{o}dinger equation
\begin{equation}
    - \dv[2]{\psi}{x} + \frac{1}{2g}\text{sn}^2(\sqrt{g}x,m) \psi = E \psi\,,
\end{equation}
by the identifications
\begin{equation}
    g \rightarrow \frac{1}{\sqrt{2 m \nu (\nu+1)}} \, , \quad E \rightarrow \frac{h}{\sqrt{2 m \nu (\nu+1)}}.
\end{equation}
Therefore, the Lam\'{e} equation can be studied as a quantum mechanical system with the potential $V_{\text{Lam\'e}}(x,m) = 1/2\,\text{sn}^2(x,m)$. For the Lam\'e potential, the instanton action and the residue then read
\begin{equation}
    S_\text{Lam\'e}(m) = \frac{2 \tanh^{-1}(\sqrt{m})}{\sqrt{m}}\,, \quad R_\text{Lam\'e}(m) = \frac{i}{\sqrt{m}},
\end{equation}
and one obtains
\begin{equation} \label{PNP-SN}
\frac{\partial E_\text{Lam\'e}}{\partial B}= -g \left[ \frac{1+m}{2} B + \frac{1-m}{2}  g \frac{\partial A_\text{Lam\'e}}{\partial g} \right.\\ \left. - \left(1-m \right) m \frac{\partial A_\text{Lam\'e}}{\partial m} \right] \, .
\end{equation}

The energy levels and the instanton function for the Lam\'{e} potential are given by

\begin{widetext}
\begin{align}
    E_{\text{Lam\'e}}(B, g; m) &= B - g \left( \frac{1}{16} + \frac{m}{16} +B^2 \left(\frac{1}{4}+\frac{m}{4}\right) \right) - g^2 \left(B \left(\frac{3}{64}-\frac{7 m}{32}+\frac{3 m^2}{64}\right) + B^3 \left(\frac{1}{16}-\frac{m}{8}+\frac{m^2}{16}\right)\right) \nonumber \\
    &- g^3 \left( \frac{9}{2048} +\frac{17}{256}B^2  + \frac{5}{128}B^4\right) \left( 1-m-m^2+m^3\right) + \mathcal{O}(g^4) \\
    A_{\text{Lam\'e}}(B, g; m) &= \frac{S_\text{Lam\'e}(m)}{g} + g \left( \frac{3}{32} +\frac{3 m}{32}+\frac{1}{4} \sqrt{m} \tanh ^{-1}\left(\sqrt{m}\right) +B^2 \left(\frac{3}{8}+\frac{3 m}{8}\right) \right) \nonumber \\
    &+ g^2 \left(B \left(\frac{17}{128}-\frac{41 m}{192}+\frac{17 m^2}{128}\right)+ B^3 \left(\frac{5}{32}-\frac{m}{48}+\frac{5 m^2}{32}\right)\right) + \mathcal{O}(g^3) \, ,
\end{align}
and they indeed satisfy Eq.~(\ref{PNP-SN}).
\end{widetext}

\bibliography{refs.bib}

\end{document}